\documentclass[prd,aps,showpacs,nofootinbib,tightenlines]{revtex4}
\usepackage[colorlinks,urlcolor=blue,linkcolor=blue,citecolor=blue]{hyperref} %
\usepackage{epsfig,graphicx,amsmath,amssymb,float,color,xspace}
\begin{document}
\title{\large {Contributions of the kaon pair from $\rho(770)$ for the three-body decays $B \to D K\bar{K}$}}

\author{Ai-Jun Ma$^1$}        \email{theoma@163.com}
\author{Wen-Fei Wang$^2$} \email{wfwang@sxu.edu.cn}  

\affiliation{$^1$Department of Mathematics and Physics,  Nanjing Institute of Technology, Nanjing, Jiangsu 211167, P.R. China \\
                 $^2$Institute of Theoretical Physics, Shanxi University, Taiyuan, Shanxi 030006, P.R. China  }
\date{\today}
\begin{abstract}
We study the contributions of the kaon pair originating from the resonance $\rho(770)$ for the three-body decays $B \to D K\bar{K}$
by employing the perturbative QCD approach. According to the predictions in this work, the contributions from the intermediate state
$\rho(770)^0 $ are relatively small for the three-body decays such as $B^0 \to \bar{D}^0 K^+ K^-$,
$B_s^0 \to \bar{D}^0 K^+ K^-$, and $B^+ \to D_s^+ K^+K^-$, while about $20\%$ of the total three-body
branching fraction for $B^+ \to \bar{D}^0 K^+ \bar{K}^0$ could possibly come from the subprocess $\rho(770)^+\to K^+ \bar{K}^0$.
We also estimate the branching fractions for $\rho(770)^\pm$ decay into the kaon pair to be
about $1\%$, and that for the neutral $\rho(770)$ decay into $K^+K^-$ or $K^0\bar{K}^0$ to be about $0.5\%$, which will be tested
by future experiments.
\end{abstract}
\maketitle

\section{INTRODUCTION}\label{sec-int}

Three-body hadronic $B$ meson decays are much more complicated than the two-body cases partly due to the entangled
resonant and nonresonant contributions, but these decay processes provide us many advantages for the study of
spectroscopy, the testing of factorization and the extraction of the CKM angles from the $CP$ asymmetries~\cite{FCPC2016-007}.
Attempts have been made to describe the whole region of the Dalitz plot for the three-body $B$ decays~\cite{1512.09284,
npb899-247,jhep1710-117}, but more attention has been focused on the resonance contributions originating from the low-energy
scalar, vector, and tensor intermediate states in the subprocesses of the three-body hadronic $B$-meson decays
 within different methods, such as the QCD factorization (QCDF)~\cite{plb622-207,prd74-114009,
prd79-094005,prd81-094033,plb699-102,prd72-094003,prd76-094006,prd88-114014,prd89-074025,prd94-094015,2005.06080,
prd102-053006,prd89-094007,epjc75-536,epjc78-845,prd99-076010} and the perturbative QCD (PQCD) approach~\cite{plb561-258,
prd91-094024,epjc76-675,plb763-29,prd95-056008,prd96-036014,epjc77-199,prd97-03306,prd98-056019,prd98-113003,epjc79-39,
epjc79-792,jhep2003-162,prd101-016015,epjc80-394,epjc80-517,epjc80-815}. In addition, there are many works within the
symmetries one can find in Refs.~\cite{plb564-90,prd72-075013,prd72-094031,prd84-056002,plb726-337,plb727-136,prd89-074043,
plb728-579,prd91-014029,prd84-034040,prd89-094013,prd92-054010} dedicated to the relevant decay modes.

The decays of the $B$ meson into a charmed $D$ meson plus kaon pair, offering rich opportunities to study the resonant
components in the $DK$ or $KK$ system, have been measured in the past two decades~\cite{plb542-171,prl100-171803,prd91-032008,
prl109-131801,jhep1801-131,prd98-072006}. The analysis of the $B \to D^{(*)}K^-K^{0(*)}$ decays was performed for the
first time by the Belle Collaboration with the detailed investigation of the invariant mass and the polarization distributions
of the $K^-K^{0(*)}$ pair~\cite{plb542-171}. In Ref.~\cite{prl100-171803}, the BaBar Collaboration reported their measurement
for the process $B^- \to D_s^+K^-K^-$. In the later study~\cite{prd91-032008} by Belle, a significant deviation
from the simple phase-space model in the $D_sK$ invariant mass distribution was found. In the recent works by the LHCb Collaboration,
observations of the decays $B^0 \to \bar{D}^0 K^+ K^-$~\cite{prl109-131801}, $B_s^0 \to \bar{D}^0 K^+ K^-$~\cite{prd98-072006},
and $B^{+} \to D_s^{+}K^+K^-$~\cite{jhep1801-131}, together with the measurements of corresponding branching fractions, were presented.
 Moreover, the studies on the $B \to D\phi(1020)$ decays, where the $\phi(1020)$ meson was reconstructed through its decay to
a $K^+K^-$ pair, were performed in Refs.~\cite{jhep1801-131,plb319-365,prd73-011103,jhep1302-043,plb727-403,prd98-071103}
by the CLEO, BaBar, and LHCb Collaborations.

The vector state $\phi(1020)$ in the $K\bar{K}$ invariant-mass spectrum for the three-body hadronic $B$-meson decays
has attracted much attention~\cite{jhep1801-131,prd85-112010,prd71-092003,JHEP1708-037,prl123-231802},
but one should note that the $P$-wave resonance contributions of the kaon pair can
also come from $\rho(770)$, $\omega(782)$ and their excited states~\cite{epjc39-41,prd81-094014,prd96-113003}.
Besides these, the charged $\rho(770)$ and its excited states are the only possible sources of vector intermediate states for
the $K^+\bar{K}^0$ or $K^-K^0$ system in the three-body $B$ decays. Although the pole mass of $\rho(770)$ is below the
threshold of the kaon pair, the virtual contribution~\cite{plb791-342,prd94-072001,prd90-072003} from the Breit-Wigner
(BW)~\cite{BW-model} tail of $\rho(770)$ for the $K\bar{K}$ was found indispensable for specific processes, such as
$\pi^-p(n)\to K^-K^+n(p)$~\cite{prd15-3196,prd22-2595}, $e^+e^- \to K^+K^-$~\cite{plb669-217,prd88-032013,prd94-112006,
plb779-64,prd99-032001}, and $\pi \pi \to K\bar K$ scattering~\cite{epjc78-897}. Recently, the component $\rho(1450)^0\to K^+K^-$
in the decays $B^\pm\to \pi^\pm K^+K^-$ was reported by LHCb to be $30\%$ of the total fit fraction and much larger than
the fit fraction $0.3\%$ from $\phi(1020)$~\cite{prl123-231802}. The subprocess $\rho(1450)^0\to K^+K^-$ and the related topics
for the decays $B^\pm\to \pi^\pm K^+K^-$ have been analyzed in Refs.~\cite{prd101-111901,prd102-053006,2007-13141} recently,
and the contribution in these decays for $K^+K^-$ from $\rho(770)^0$, which has been ignored in the experimental and theoretical
studies, was found to be of the same order of that from $\rho(1450)^0$ in Ref.~\cite{prd101-111901}.

In the previous works~\cite{npb923-54,prd96-093011,epjc79-539,plb788-468,plb791-342,prd100-014017,cpc43-073103,
prd102-056017,ijmpa35-2050164}, the resonance contributions from various intermediate states for the three-body decays
$B \to Dh_1h_2$ ($h_{1,2}$ stands for pion or kaon) have been studied within the PQCD approach based on the $k_{T}$
factorization theorem~\cite{plb504-6,prd63-054008,prd63-074009,ppnp51-85}. In this work, we shall focus on the contributions
of the subprocesses $\rho(770)\to K\bar K$ for the three-body decays $B \to D K\bar{K}$, where the symbol $\bar{K}$ means
the kaons $K^+$ and $K^0$, and the symbol $K$ means the kaons $K^-$ and $\bar{K}^0$. In view of the narrow decay width
of $\omega(782)$ and the gap between its pole mass and the threshold of the kaon pair, the branching fractions for the decays
with the subprocess $\omega(782) \to K\bar{K}$ are small and negligible compared with the contribution from
$\rho(770) \to K\bar{K}$ in the same decay mode~\cite{prd101-111901}.  Meanwhile, there are still disparities between the fitted
coefficients of the timelike form factors for kaons from currently known experimental results~\cite{epjc39-41,prd81-094014,
jetp129-386}, we will leave the possible subprocesses with those excited states of $\rho(770)$ and $\omega(782)$ decay into
$K\bar K$ to future study.

The rest of this paper is organized as follows: In Sec.~\ref{sec-fra}, we give a brief review of the PQCD framework for
the concerned decay processes. The numerical results and the phenomenological analyses are given in Sec.~\ref{sec-res}.
The summary of this work is presented in Sec.~\ref{sec-con}. The relevant quasi-two-body decay amplitudes are collected
in the Appendix.

\section{FRAMEWORK} \label{sec-fra}

\begin{figure}[tbp]
\begin{center}
\centerline{\epsfxsize=15cm \epsffile{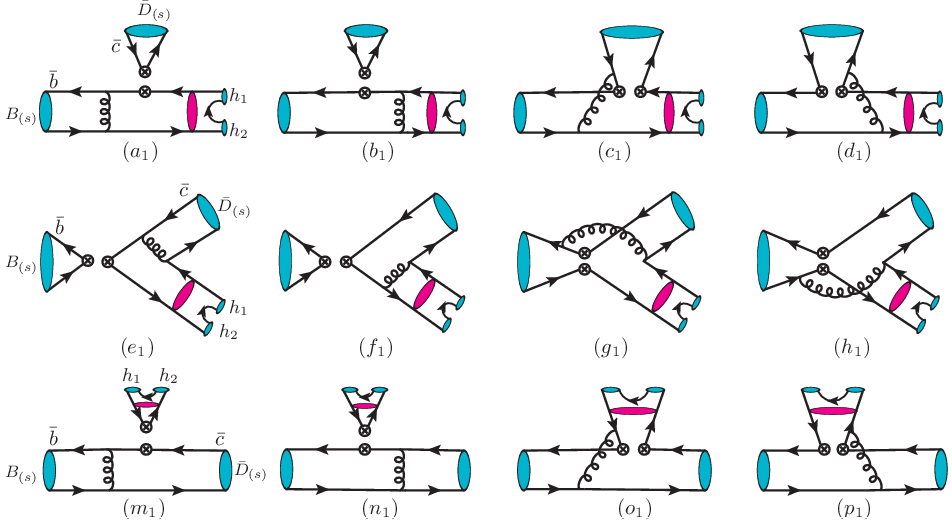}}
\caption{The leading-order Feynman diagrams for the quasi-two-body decays $B_{(s)} \to \bar{D}_{(s)} \rho(770) \to \bar{D}_{(s)}K \bar{K}$.
 The label $h_1h_2$ denotes the kaon pair, and the pink ellipse represents the intermediate state $\rho(770)$. }
\label{fig:fig1}
\end{center}
\vspace{-0.8cm}
\end{figure}
\begin{figure}[tbp]
\begin{center}
\centerline{\epsfxsize=15cm \epsffile{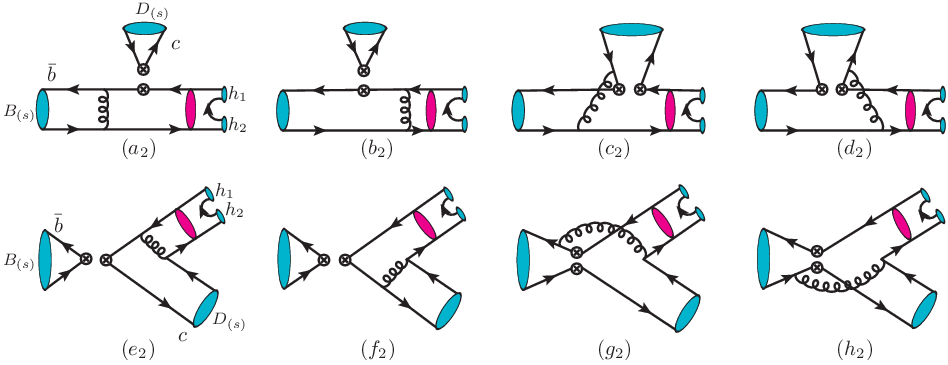}}
\caption{ The leading-order Feynman diagrams for the quasi-two-body decays $B_{(s)} \to  D_{(s)}  \rho(770) \to  D_{(s)} K \bar{K}$.
The label $h_1h_2$ denotes the kaon pair, and the pink ellipse represents the intermediate state $\rho(770)$.}
\label{fig:fig2}
\end{center}
\vspace{-0.8cm}
\end{figure}

In the light-cone coordinates, the momenta $p_{B}$,  $p$, and $p_3$ for the $B$ meson, the resonance $\rho$, and the final state
$D$, respectively, are chosen as
\begin{eqnarray}\label{mom-pBpp3}
p_{B}=\frac{m_{B}}{\sqrt2}(1,1,0_{\rm T}),~\quad p=\frac{m_{B}}{\sqrt2}(1-r^2,\eta,0_{\rm T}),~\quad
p_3=\frac{m_{B}}{\sqrt2}(r^2,1-\eta,0_{\rm T}),
\end{eqnarray}
where $m_{B}$ denotes the mass for the $B$ meson, the variable $\eta$ is defined as $\eta=s/(m^2_{B}-m^2_D)$,
the invariant mass square $s=p^2=m^2_{K\bar K}$ for the kaon pair, and the mass ratio $r=m_{D}/m_{B}$.
The momenta of the light quark in the $B$ meson, $\rho$, and the $D$ meson are denoted as $k_B$, $k$, and $k_3$, with
\begin{eqnarray}
k_B=(0,x_B\frac{m_B}{\sqrt{2}},{k}_{B{\rm T}}),~\quad
k=(z\frac{(1-r^2)m_B}{\sqrt{2}},0,{k}_{\rm T}),~\quad
k_3=(0,x_3\frac{(1-\eta)m_B}{\sqrt{2}},{k}_{3{\rm T}}),
 \end{eqnarray}
where the momentum fractions $x_{B}$, $z$, and $x_3$ run between zero and unity.

The decay amplitude ${\mathcal A}$ for the quasi-two-body processes  $B\to D \rho(770) \to D K\bar{K}$ in the PQCD approach
can be expressed as the convolution of a hard kernel $H$ containing one hard gluon exchange with the relevant hadron distribution
amplitudes~\cite{plb561-258,prd89-074031}
\begin{eqnarray} \label{def-DA}
{\cal A}=\Phi_B\otimes H\otimes \Phi_{D}\otimes\Phi_{KK},
\end{eqnarray}
where the distribution amplitudes $\Phi_B$, $\Phi_{D}$, and $\Phi_{KK}$ for the initial- and final-state mesons absorb the
nonperturbative dynamics. In this work, we employ the same distribution amplitudes for $B$ and $D$ mesons as those widely
adopted in the studies of the hadronic $B$-meson decays in the PQCD approach; one can find their explicit expressions and
parameters in Ref.~\cite{npb923-54} and the references therein.

The $P$-wave $K\bar{K}$ system distribution amplitudes along with the subprocesses $\rho(770)\to K\bar{K}$
are defined as~\cite{prd101-111901,epjc80-815}
\begin{eqnarray}\label{def-DA-KK}
     \phi^{P\text{-wave}}_{KK}(z,s)
             =\frac{-1}{\sqrt{2N_c}} \left[\sqrt{s}\,{\epsilon\hspace{-1.5truemm}/}\!_L\phi^0(z,s)
                      +  {\epsilon\hspace{-1.5truemm}/}\!_L {p\hspace{-1.7truemm}/} \phi^t(z,s)+\sqrt s \phi^s(z,s)  \right],
\end{eqnarray}
where $z$ is the momentum fraction for the spectator quark, $s$ is the squared invariant mass of the kaon pair, and $\epsilon_L$ and
$p$ are the longitudinal polarization vector and momentum for the resonance. The twist-$2$ and twist-$3$ distribution amplitudes
$\phi^{0}$, $\phi^{s}$, and $\phi^{t}$ are parameterized as~\cite{prd101-111901}
\begin{eqnarray}
   \phi^{0}(z,s)&=&\frac{3F^\rho_K(s)}{\sqrt{aN_c}} z(1-z)\left[1+a_2^{0} C^{3/2}_2(1-2z) \right]\!,\label{def-DA-0}\\
  \phi^{s}(z,s)&=&\frac{3F^s_K(s)}{2\sqrt{aN_c}}(1-2z)\left[1+a_2^s\left(1-10z+10z^2\right) \right]\!,\label{def-DA-s}\\
   \phi^{t}(z,s)&=&\frac{3F^t_K(s)}{2\sqrt{aN_c}}(1-2z)^2\left[1+a_2^t  C^{3/2}_2(1-2z)\right]\!,\label{def-DA-t}\quad
\end{eqnarray}
with the Gegenbauer polynomial $C^{3/2}_2(t)=3\left(5t^2-1\right)/2$,
$F^{s,t}_K(s)\approx (f^T_{\rho}/f_{\rho})F^{\rho}_K(s)$~\cite{plb763-29}, $a=1$ for $\rho(770)^0$, and $a=2$ for $\rho(770)^\pm$.
In the numerical calculation, we adopt
$f_\rho=0.216$ GeV ~\cite{prd75-054004,jhep1608-098}  and $f^T_\rho=0.184$ GeV~\cite{prd80-054510}.
The Gegenbauer moments $a_2^{0,s,t}$ are the same as those in the distribution amplitudes for the intermediate state $\rho(770)$
 in Refs.~\cite{plb763-29,prd101-111901}. The vector timelike form factors for kaons are written as~\cite{prd88-114014}
\begin{eqnarray}
  F_{K^+K^-}^{u}&=&F_{K^0\bar K^0}^{d}= F_\rho+3 F_\omega, \\
  F_{K^+K^-}^{d}&=&F_{K^0\bar K^0}^{u}=-F_\rho+3 F_\omega, \\
  F_{K^+K^-}^{s}&=&F_{K^0\bar K^0}^{s}=-3 F_\phi,
  \label{def-F-uds}
\end{eqnarray}
where $F_\rho$,  $F_\omega$, and $F_\phi$ come from  the definitions of the electromagnetic form factors for the charged and
neutral kaon~\cite{epjc39-41,prd81-094014}:
\begin{eqnarray}
  F_{K^+}^{I=1}(s)&=&+\frac12\sum_\rho c^K_\rho {\rm BW}_\rho(s) +\frac16\sum_\omega c^K_\omega {\rm BW}_\omega(s)
                   +\frac13\sum_\phi c^K_\phi {\rm BW}_\phi(s)=F_\rho+F_\omega+F_\phi,
  \label{def-F-K+}   \\
  F_{K^0}^{I=1}(s)&=&-\frac12\sum_\rho c^K_\rho {\rm BW}_\rho(s) +\frac16\sum_\omega c^K_\omega {\rm BW}_\omega(s)
                   +\frac13\sum_\phi c^K_\phi {\rm BW}_\phi(s)=-F_\rho+F_\omega+F_\phi.
  \label{def-F-K0}
\end{eqnarray}
The symbol $\sum$ means the summation for the resonances $\rho(770)$, $\omega(782)$, or $\phi(1020)$ and their
corresponding excited states. The normalization factors $c^K_V$ for resonances determined by fitting experimental
data and the corresponding BW formula can be found in Refs.~\cite{epjc39-41,prd81-094014,jetp129-386}.
It is not difficult to find that the corresponding coefficients $c^K_V$ for $\rho(770)$, $\omega(782)$, or $\phi(1020)$ are
close to each other in Refs.~\cite{epjc39-41,prd81-094014,jetp129-386}, while it can be shown that
those coefficients for the excited states have significant
differences by comparing the fitted parameters in  Table~2 in Refs.~\cite{epjc39-41,prd81-094014} and Table~1
in Ref.~\cite{jetp129-386}. In this work, we concern ourselves only with the $\rho(770)$ components of
the vector kaon timelike form factors;
the fitted values for the coefficients $C^K_{\rho(770)}$  in the kaon form factors collected from
Refs.~\cite{epjc39-41,prd81-094014,jetp129-386} have been listed in the Table~\ref{tab0}.
The columns ``Fit(1)," ``Fit(2)" and ``Model I," ``Model II" represent the values parameterized with different constraints in each work.
Due to the closeness of the coefficients $C^K_{\rho(770)}$ in Refs.~\cite{epjc39-41,prd81-094014,jetp129-386}, we choose the
value of ``Fit(1)" in Ref.~\cite{epjc39-41} in our numerical calculation. The resonance shape for $\rho(770)$ is described by the
KS version of the BW formula~\cite{zpc48-455,epjc39-41}:
\begin{eqnarray}\label{def-BW}
 \frac{m_{\rho}^2}{m_\rho^2-s-i\sqrt{s}\Gamma_{tot}(s)}\,,    \label{BW}
\end{eqnarray}
 where the effective $s$-dependent width is given by
 \begin{eqnarray}\label{def-width}
 \Gamma_{tot}(s)\approx \Gamma_{\rho \to 2\pi}(s)=\Gamma_\rho\frac{m_\rho^2}{s}
\left(\frac{\beta(s,m_\pi)}{\beta(m_\rho^2,m_\pi)}\right)^3
\end{eqnarray}
with $\beta(s,m)=\sqrt{1-4m^2/s}$. In addition, one has the timelike form factor for $K^+\bar{K}^0$ and $K^-K^0$ from
the relation~\cite{prd96-113003,epjc39-41}
\begin{eqnarray}
 F_{K^+\bar{K}^0}(s)=-F_{K^-K^0}(s)=2F_{K^+}^{I=1}(s)
  \label{def-F-K+0}
 \end{eqnarray}
and should keep only the $\rho$ resonance contributions with isospin symmetry.

\begin{table}[htb]
\begin{center}
\caption{The fitted results for the coefficients $C^K_{\rho(770)}$ of the kaon form factors.}
\label{tab0}   
\begin{tabular}{ l | c c|c c| c c  }
 \hline\hline
  ~         & {\rm Fit(1)}~\cite{epjc39-41}         & {\rm Fit(2)}~\cite{epjc39-41}
             & {\rm Fit(1)}~\cite{prd81-094014}   & {\rm Fit(2)}~\cite{prd81-094014}
              & {\rm Model I}~\cite{jetp129-386}  & {\rm Model II}~\cite{jetp129-386} \\
 \hline
 $C^K_{\rho(770)}$\; & $1.195 \pm 0.009$  & $1.139 \pm 0.010$
                                  & $1.138 \pm 0.011$  & $1.120 \pm 0.007$
                                  & $1.162 \pm 0.005$  & $1.067 \pm 0.041$ \\
 \hline\hline
\end{tabular}
\end{center}
\end{table}

For the decays $B_{(s)} \to  \bar{D}_{(s)} \rho(770) \to \bar{D}_{(s)} K \bar{K}$ and the CKM-suppressed decays
$B_{(s)} \to  D_{(s)} \rho(770) \to D_{(s)} K \bar{K}$, the effective Hamiltonian ${\cal  H}_{eff}$ can be expressed as
\begin{eqnarray}\label{def-heff}
{\cal  H}_{eff}&=& \left\{
  \begin{array}{ll}
      \frac{G_F}{\sqrt{2}}V^*_{cb}V_{ud(s)} \left[ C_1(\mu)O_1(\mu)+C_2(\mu)O_2(\mu) \right],
             & \ \  {\rm for} \ \ B_{(s)} \to  \bar{D}_{(s)} \rho(770) \to \bar{D}_{(s)} K \bar{K}\ \ {\rm decays},\\
      \frac{G_F}{\sqrt{2}} V^*_{ub}V_{cd(s)} \left[ C_1(\mu)O_1(\mu)+C_2(\mu)O_2(\mu) \right],
            & \ \  {\rm for} \ \ B_{(s)} \to  D_{(s)} \rho(770) \to D_{(s)} K \bar{K} \ \ {\rm decays},\\
  \end{array} \right.
\end{eqnarray}
where $G_F = 1.16638 \times 10^{-5}~{\rm GeV}^{-2}$, $V_{ij}$ are the CKM matrix elements, $C_{1,2}(\mu)$ represent
the Wilson coefficients  at the renormalization scale $\mu$, and $O_{1,2}$ are the local four-quark operators. According to
the typical Feynman diagrams for the concerned decays as shown in Figs.~\ref{fig:fig1} and~\ref{fig:fig2}, the decay amplitudes
for  $B_{(s)} \to \bar{D}_{(s)} \rho(770)$  with the  subprocesses $\rho(770)^0 \to K^+K^-/K^0\bar{K}^0$ and
$\rho(770)^+ \to K^+\bar{K}^0$ are given as follows:
\begin{eqnarray}\label{def-DA-Dbar}
\mathcal{A}({B^+ \to {\bar{D^0}} \rho^+})&=&\frac{G_F}{\sqrt2}V^*_{cb}V_{ud} [a_1F_{e\rho}^{LL}+C_2
M_{e\rho}^{LL}+a_2F_{eD}^{LL}+C_1M_{eD}^{LL} ], \\
\mathcal{A}({B^0 \to {D^-} \rho^+   })&=&\frac{G_F}{\sqrt2}V^*_{cb}V_{ud} [a_1F_{a\rho}^{LL}+C_2
M_{a\rho}^{LL}+a_2F_{eD}^{LL}+C_1M_{eD}^{LL} ], \\
\mathcal{A}({B^0 \to {\bar{D^0}} \rho^0 })&=&\frac{G_F}{2} V^*_{cb}V_{ud}
[a_1(-F_{e\rho}^{LL}+F_{a\rho}^{LL})+C_2(-M_{e\rho}^{LL}+M_{a\rho}^{LL}) ],\\
\mathcal{A}({B_s^0 \to {D^-} \rho^+})&=&\frac{G_F}{\sqrt2}V^*_{cb}V_{us} [a_1F_{a\rho}^{LL}+C_2
M_{a\rho}^{LL} ], \\
\mathcal{A}({B_s^0 \to {\bar{D^0}} \rho^0   })&=&\frac{G_F}{2}V^*_{cb}V_{us} [a_1F_{a\rho}^{LL}+C_2
M_{a\rho}^{LL} ], \\
\mathcal{A}({B_s^0 \to {D_s^-} \rho^+ })&=&\frac{G_F}{\sqrt2}V^*_{cb}V_{ud} [a_2F_{eD}^{LL}+C_1
M_{eD}^{LL} ],
\end{eqnarray}
while the decay amplitudes for $B_{(s)} \to D_{(s)} \rho(770)$  with the  subprocesses
$\rho(770)^0 \to K^+K^-/K^0\bar{K}^0$, $\rho(770)^+ \to K^+\bar{K}^0$ and  $\rho(770)^- \to K^-K^0$  can be written as
\begin{eqnarray}\label{def-DA-D}
\mathcal{A}({B^+ \to {D^0} \rho^+  })&=&\frac{G_F}{\sqrt2}V^*_{ub}V_{cd}
[a_1F_{e\rho}^{LL}+C_2M_{e\rho}^{LL}+a_2F_{aD}^{LL}+C_1M_{aD}^{LL} ], \\
\mathcal{A}({B^+ \to {D^+} \rho^0  })&=&\frac{G_F}{2} V^*_{ub}V_{cd} [a_2(F_{e\rho}^{LL}-F_{aD}^{LL})
+C_1(M_{e\rho}^{LL}-M_{aD}^{LL})], \\
\mathcal{A}({B^+ \to {D_s^+} \rho^0  })&=&
\frac{G_F}{2}V^*_{ub}V_{cs} [a_2F_{e\rho}^{LL}+C_1M_{e\rho}^{LL} ],\\
\mathcal{A}({B^0 \to {D^0} \rho^0  })&=&\frac{G_F}{2} V^*_{ub}V_{cd}
[a_1(-F_{e\rho}^{LL}+F_{aD}^{LL})+C_2(-M_{e\rho}^{LL}+M_{aD}^{LL}) ],\\
\mathcal{A}({B^0 \to {D^+} \rho^-  })&=&\frac{G_F}{\sqrt2}V^*_{ub}V_{cd} [a_2F_{e\rho}^{LL}+C_1
M_{e\rho}^{LL}+a_1F_{aD}^{LL}+C_2M_{aD}^{LL} ],\\
\mathcal{A}({B^0 \to {D_s^+} \rho^-  })&=&\frac{G_F}{\sqrt2}V^*_{ub}V_{cs}
[a_2F_{e\rho}^{LL}+C_1M_{e\rho}^{LL} ],\\
\mathcal{A}({B_s^0 \to {D^0} \rho^0  })&=&\frac{G_F}{2}V^*_{ub}V_{cs} [a_1F_{aD}^{LL}+C_2M_{aD}^{LL} ],\\
\mathcal{A}({B_s^0 \to {D^+} \rho^-  })&=&\frac{G_F}{\sqrt2}V^*_{ub}V_{cs}
[a_1F_{aD}^{LL}+C_2M_{aD}^{LL} ],
\end{eqnarray}
with the Wilson coefficients $a_1=C_1+ C_2/3$ and $a_2=C_2+ C_1/3$. The explicit expressions of individual amplitude
$F$ and $M$ for the factorizable and nonfactorizable Feynman diagrams can be found in Appendix.

The differential branching fractions ($\mathcal B$) for the quasi-two-body decays $B\to D \rho(770) \to DK\bar K$ can be written
as~\cite{prd79-094005,prd101-111901,epjc80-815}
\begin{eqnarray}
 \frac{d{\mathcal B}}{d\eta}=\tau_B\frac{q^3 q^3_D}{12\pi^3m^5_B}|{\mathcal A}|^2\;.
\label{eqn-diff-bra}
\end{eqnarray}
The magnitudes of the momenta for $K$ and $D$ in the center-of-mass frame of the kaon pair are written as
\begin{eqnarray}
       q&=&\frac{1}{2}\sqrt{s-4m^2_K},  \label{def-q}\\
   q_D&=&\frac{1}{2\sqrt s}\sqrt{\left(m^2_{B}-m_{D}^2\right)^2 -2\left(m^2_{B}+m_{D}^2\right)s+s^2}.   \label{def-qh}
\end{eqnarray}

\section{RESULTS} \label{sec-res}

In the numerical calculations, the input parameters, such as masses and decay constants (in units of GeV)
and $B$-meson lifetimes (in units of ps), are adopted as follows~\cite{PDG2020}:
\begin{eqnarray}
 m_{B^\pm}&=&5.279,\quad m_{B^0}=5.280,\quad m_{B_s^0}=5.367,\quad m_{D^\pm}=1.870, \quad m_{D^0}=1.865, \nonumber\\
 m_{D_s^\pm}&=&1.968, \quad m_{K^\pm}=0.494,\quad m_{K^0}=0.498, \quad m_{c}=1.27,\qquad m_{\pi^\pm}=0.140,\nonumber\\
 m_{\pi^0}&=&0.135,\quad f_{B}=0.189,\qquad f_{B_s}=0.231, \quad\; f_{D}=0.2126, \quad\;\; f_{D_s}=0.2499, \nonumber\\
 \tau_{B^\pm}&=& 1.638,\quad \tau_{B^0}= 1.519, \quad~\; \tau_{B_s^0}= 1.515. \label{eq:inputs}
\end{eqnarray}
For the Wolfenstein parameters of the CKM mixing matrix, we use the values
$A=0.790^{+0.017}_{-0.012},~\lambda=0.22650\pm0.00048$, $\bar{\rho} = 0.141^{+0.016}_{-0.017}$, and $\bar{\eta}= 0.357\pm0.011$
as listed in Ref.~\cite{PDG2020}.

In Tables~\ref{tab1} and \ref{tab2}, we list our numerical results for the branching fractions of
the $B_{(s)} \to  \bar{D}_{(s)}   \rho(770) \to  \bar{D}_{(s)}  K \bar{K}$ decays and
the CKM-suppressed $B_{(s)} \to  D_{(s)}   \rho(770) \to  D_{(s)} K \bar{K}$ decays. The first error of these branching fractions
comes from  the uncertainty of the $B_{(s)}$ meson shape parameter $\omega_B = 0.40 \pm 0.04$ or
$\omega_{B_s}=0.50 \pm 0.05$; the second error is induced by the uncertainties of the Gegenbauer moments $a^0_2=0.25\pm0.10$, $a^s_2=0.75\pm0.25$, and $a^t_2= -0.60\pm0.20$ in the kaon-kaon distribution amplitudes; the last one is due to
$C_{D}=0.5\pm 0.1$ or $ C_{D_s}=0.4\pm0.1$  for the $D_{(s)}$ meson wave function. The errors that come from the uncertainties
of other parameters are small and have been neglected. Since the concerned decay modes occur only through
the tree-level quark diagrams, there are no direct $CP$ asymmetries for these decays in the standard model.

\begin{table}[htb]
\begin{center}
\caption{The PQCD predictions of the branching fractions for the
$B_{(s)} \to  \bar{D}_{(s)}   \rho(770) \to  \bar{D}_{(s)}  K \bar{K}$ decays. The decay mode with the subprocess
$\rho(770)^0\to K^0\bar K^0$ has the same branching fraction  of its corresponding mode with $\rho(770)^0 \to K^+K^-$.}
\label{tab1}
\begin{tabular}{ c   c   c   } \hline\hline
{\rm   Decay modes} & {\rm   Unit}& {\rm   Quasi-two-body results} \\ \hline
$B^+ \to \bar{D}^0 \rho(770)^+ \to \bar{D}^0 K^+ \bar{K}^0$
&\;\;\;$(10^{-4})$\;\;\;&$1.18^{+0.62}_{-0.40}(\omega_B)^{+0.09}_{-0.12}(a^0_2+a^s_2+a^t_2)^{+0.07}_{-0.09}(C_D)$  \\
$B^0 \to D^- \rho(770)^+ \to D^- K^+ \bar{K}^0 $
&$(10^{-5})$&$7.93^{+5.01}_{-2.93}(\omega_B)^{+0.32}_{-0.30}(a^0_2+a^s_2+a^t_2)^{+0.65}_{-0.63}(C_D)$\\
$B^0 \to \bar{D}^0 \rho(770)^0 \to \bar{D}^0 K^+ K^- $ &$(10^{-6})$
&$1.07^{+0.46}_{-0.37}(\omega_B)^{+0.80}_{-0.58}(a^0_2+a^s_2+a^t_2)^{+0.01}_{-0.01}(C_D)$\\
$ B_s^0 \to D^- \rho(770)^+ \to D^- K^+ \bar{K}^0 $ &$(10^{-8})$
&$4.22^{+0.58}_{-0.67}(\omega_B)^{+0.90}_{-0.65}(a^0_2+a^s_2+a^t_2)^{+0.40}_{-0.30}(C_D)$\\
$ B_s^0 \to \bar{D}^0 \rho(770)^0 \to \bar{D}^0 K^+ K^- $ &$(10^{-8})$
&$1.05^{+0.15}_{-0.17}(\omega_B)^{+0.23}_{-0.15}(a^0_2+a^s_2+a^t_2)^{+0.10}_{-0.07}(C_D)$\\
$ B_s^0 \to D_s^- \rho(770)^+ \to D_s^-K^+ \bar{K}^0  $ &$(10^{-5})$
&$6.06^{+3.47}_{-2.06}(\omega_B)^{+0.04}_{-0.04}(a^0_2+a^s_2+a^t_2)^{+0.47}_{-0.45}(C_D)$\\
 \hline\hline
\end{tabular} \end{center}
\end{table}

\begin{table}[htb]
\begin{center}
\caption{The PQCD predictions of the branching fractions for the CKM-suppressed
$B_{(s)} \to  D_{(s)}   \rho(770) \to  D_{(s)} K \bar{K}$ decays. The decay mode with the subprocess $\rho(770)^0\to K^0\bar K^0$
has the same branching fraction of its corresponding decay with $\rho(770)^0 \to K^+K^-$.}
\label{tab2}
\begin{tabular}{ c  c c } \hline\hline
{\rm  Decay modes} & {\rm Unit}& {\rm Quasi-two-body results}  \\ \hline
$B^+ \to D^0 \rho(770)^+ \to D^0  K^+\bar{K}^0 $
&\;\;\;$(10^{-10})$\;\;\;&$5.27^{+1.23}_{-0.59}(\omega_B)^{+2.49}_{-1.68}(a^0_2+a^s_2+a^t_2)^{+0.33}_{-0.08}(C_D)$ \\
$B^+ \to D^+ \rho(770)^0 \to D^+ K^+K^- $
&$(10^{-9})$&$3.22^{+0.52}_{-0.45}(\omega_B)^{+0.86}_{-0.43}(a^0_2+a^s_2+a^t_2)^{+0.01}_{-0.01}(C_D)$ \\
$B^+ \to D_s^+ \rho(770)^0 \to D_s^+ K^+K^- $
&$(10^{-8})$&$6.26^{+1.69}_{-1.30}(\omega_B)^{+2.69}_{-0.92}(a^0_2+a^s_2+a^t_2)^{+0.03}_{-0.02}(C_D)$\\
$ B^0 \to D^0 \rho(770)^0 \to D^0K^+K^- $
&$(10^{-11})$&$7.79^{+2.02}_{-1.33}(\omega_B)^{+4.63}_{-2.86}(a^0_2+a^s_2+a^t_2)^{+0.81}_{-0.61}(C_D)$ \\
$ B^0 \to D^+ \rho(770)^- \to D^+ K^0K^- $
&$(10^{-9})$&$6.87^{+2.05}_{-1.60}(\omega_B)^{+3.30}_{-1.01}(a^0_2+a^s_2+a^t_2)^{+0.08}_{-0.08}(C_D)$\\
$ B^0 \to D_s^+ \rho(770)^- \to D_s^+ K^0K^- $
&$(10^{-7})$&$2.32^{+0.63}_{-0.48}(\omega_B)^{+1.00}_{-0.34}(a^0_2+a^s_2+a^t_2)^{+0.01}_{-0.01}(C_D)$ \\
$ B_s^0 \to D^0 \rho(770)^0 \to D^0 K^+K^-  $
&$(10^{-9})$&$1.85^{+0.36}_{-0.32}(\omega_B)^{+0.61}_{-0.45}(a^0_2+a^s_2+a^t_2)^{+0.09}_{-0.08}(C_D)$   \\
$ B_s^0 \to D^+ \rho(770)^- \to D^+ K^0K^-  $
&$(10^{-9})$&$7.47^{+1.49}_{-0.32}(\omega_B)^{+2.42}_{-1.83}(a^0_2+a^s_2+a^t_2)^{+0.40}_{-0.37}(C_D)$ \\
 \hline\hline
\end{tabular} \end{center}
\end{table}

The predictions for the branching fractions of the decays $B_{(s)} \to  D_{(s)}   \rho(770) \to  D_{(s)} K \bar{K}$ in
Table~\ref{tab1} are generally smaller than  the corresponding results for the
$B_{(s)} \to  \bar{D}_{(s)}   \rho(770) \to  \bar{D}_{(s)}  K \bar{K}$ decays in Table~\ref{tab2} due to the
strong CKM suppression factor $| \frac{V_{ub}^* V_{cd}}{V_{cb}^*V_{ud}}|^2$ or $| \frac{V_{ub}^* V_{cs}}{V_{cb}^*V_{us}}|^2$,
as discussed in Ref.~\cite{npb923-54}. The central values for the PQCD-predicted branching fractions of the decays
$B^0 \to \bar{D}^0 \rho(770)^0 \to \bar{D}^0 K^+ K^-$ and $B_{s}^0 \to \bar{D}^0 \rho(770)^0 \to \bar{D}^0 K^+ K^-$
are $0.18\%$ and $0.019\%$ of the experimental measurements
${\cal B}(B^0 \to \bar{D}^0 K^+ K^-)=(5.9\pm0.5)\times 10^{-4}$
and ${\cal B}(B_s^0 \to \bar{D}^0 K^+ K^-)=(5.5\pm0.8)\times 10^{-5}$ respectively, in the
{\it Review of Particle Physics} (Ref.~\cite{PDG2020}),
which have been averaged from the results in Refs.~\cite{prd98-072006,prl109-131801} presented by LHCb.
However, with the branching ratio $ {\cal B}{(B^+ \to \bar{D}^0 K^+ \bar{K}^0)}=(5.5\pm1.4\pm0.8)\times 10^{-4}$ presented by
the Belle Collaboration~\cite{plb542-171}, one has a sizable percent at $21.45\%$ of the total branching fraction
for the quasi-two-body decay $B^+ \to \bar{D}^0 \rho(770)^+ \to \bar{D}^0 K^+ \bar{K}^0$. This tells us that the contributions
from  $\rho(770)^{\pm} \to K\bar{K}$ could be considerably large in the relevant three-body $B$-meson decays.

In Ref.~\cite{jhep1801-131}, LHCb presented the first observation of the decay $B^+ \to D_s^+ K^+K^-$, and the branching fraction
was determined to be $(7.1\pm0.5\pm0.6\pm0.7) \times 10^{-6}$. Utilizing our prediction
${\cal B}(B^+ \to D_s^+ \rho(770)^0 \to D_s^+ K^+K^-)=(6.26^{+3.18}_{-1.59}) \times 10^{-8}$,
where the individual errors have been added in quadrature,  we obtain the ratio
$\frac{{\cal B}(B^+ \to D_s^+ \rho(770)^0 \to D_s^+ K^+K^-)}{{\cal B}(B^+ \to D_s^+ K^+K^-)}=0.88^{+0.47}_{-0.26}\%$,
which is quite small, as expected.  In addition, LHCb also gave a branching fraction for the $B^+ \to D_s^+\phi(1020)$ decay of
$(1.2^{+1.6}_{-1.4}\pm{0.8}\pm{0.1})\times 10^{-7}$ and set an upper limit as $ 4.9(4.2)\times 10^{-7}$
at the $95\%~(90\%)$ confidence level, which is roughly $1$ order smaller than their previous result in Ref.~\cite{jhep1302-043}.
By adopting ${\cal B}(\phi(1020) \to K^+K^-)=0.492$~\cite{PDG2020} and the relation between the quasi-body decay and the
corresponding two-body decay
\begin{eqnarray}
{\cal B}(B\to D R \to D h_1h_2)\approx {\cal B}(B\to D R) \cdot {\cal B}(R \to h_1h_2),
 \label{quasi-da}
\end{eqnarray}
we find that ${\cal B}(B^+ \to D_s^+ \rho(770)^0 \to D_s^+ K^+K^-)$ predicted in this work has the same magnitude as
the branching ratio for  $B^+ \to D_s^+ \phi(1020) \to D_s^+ K^+K^-$ measured by LHCb within large uncertainties,
while ${\cal B}(B^+ \to D_s^+\phi(1020) \to D_s^+ K^+K^-)$ was predicted to be $(1.53\pm0.23) \times 10^{-7}$
within the PQCD approach in Ref.~\cite{ijmpa35-2050164}.

\begin{figure}[tbp]
\centerline{\epsfxsize=8 cm \epsffile{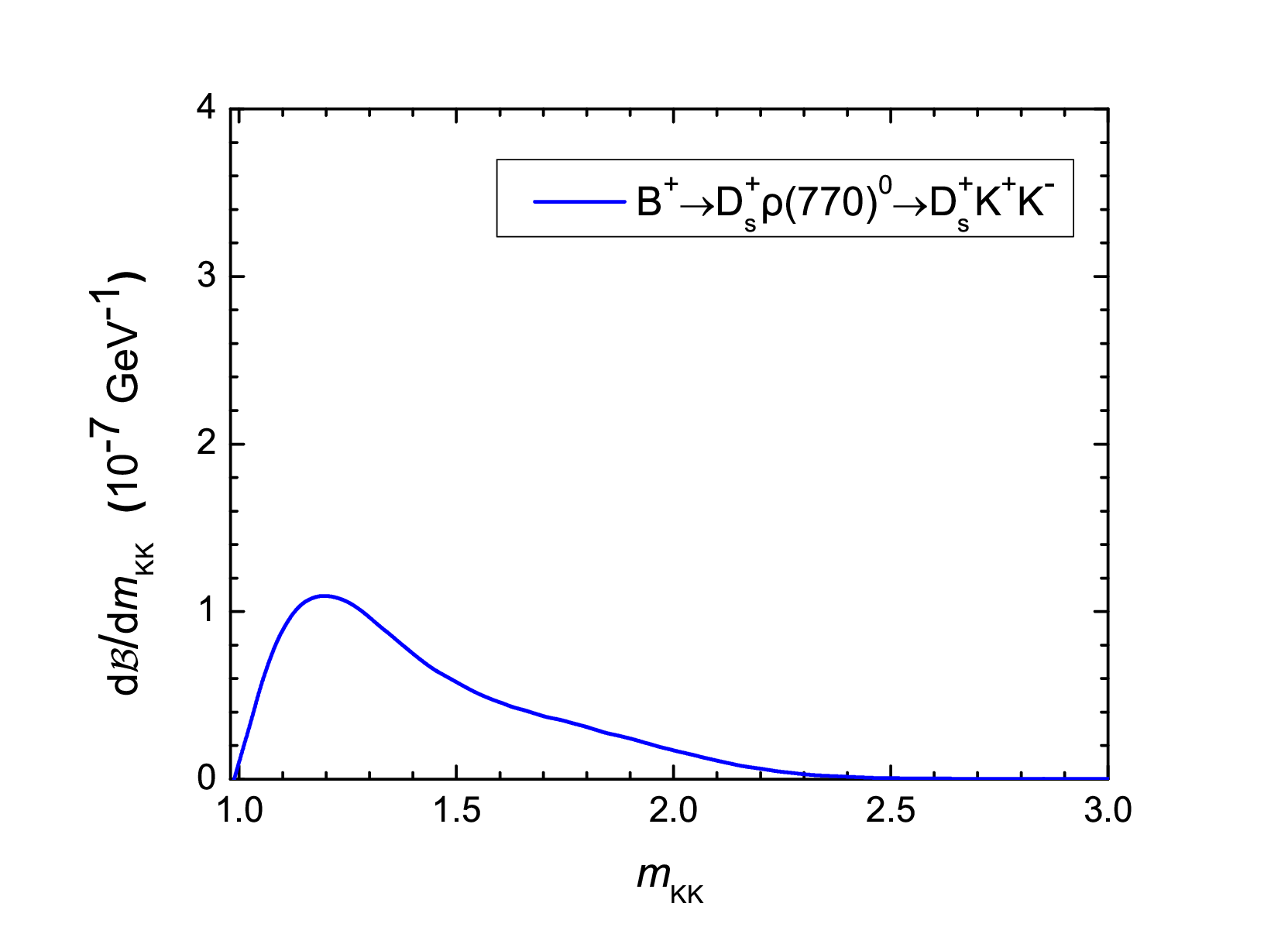} }
\vspace{-0.5cm}
\caption{The PQCD prediction for the differential branching ratio of the decay mode
               $B^+ \to D_s^+ \rho(770)^0 \to D_s^+ K^+K^-$, with the invariant mass ranges from $2m_K$ to $3~{\rm GeV}$. }
\label{fig:fig3}
\end{figure}

In Fig.~\ref{fig:fig3}, we show the  differential branching fraction of the decay mode
${\cal B}(B^+ \to D_s^+ \rho(770)^0 \to D_s^+ K^+K^-)$ with the invariant mass in the range of $[2m_K, 3~{\rm GeV}]$.
The bump in the curve is caused by the strong depression of the phase-space factors $q$ and $q_D$ in Eqs.~(\ref{def-q})
and~(\ref{def-qh}) near the $K^+K^-$ threshold. This depression near the threshold, along with the similar mass between
$K^\pm$ and $K^0,\bar{K}^0$, causes the decay channel with the subprocess $\rho(770)^0\to K^0\bar K^0$  to have the same
branching fraction as the corresponding decay mode with the subprocess $\rho(770)^0 \to K^+K^-$.

In principle, the applicability of the PQCD calculations in the high $K \bar{K}$ invariant mass region is apt to deteriorate
 because of the small energy release. Fortunately, the evolution of the kaon form factor $F_K(s)$ in the decay
 amplitude $\mathcal{A}$ will naturally suppress the resonant contribution from the region where the invariant mass of the
kaon pair is far away from the pole mass of the resonant state $\rho(770)$.
Taking the decay $B^+ \to D_s^+ \rho(770)^0 \to D_s^+ K^+K^-$ as an example, it is easy to check that the main portion of its
branching ratio lies in the region around $1.2~{\rm GeV}$, as shown in Fig.~\ref{fig:fig3}.
Numerically, the central values of its  branching ratio are calculated as $4.08 \times 10^{-8}$ and $5.85 \times 10^{-8}$
after making the integration over the ranges of $m_{K\bar{K}}$ in [$2m_K$,$1.5~{\rm GeV}$] and [$2m_K$,$2~{\rm GeV}$], respectively,
which amount to $65.18\%$ and $93.45\%$ of the
value $6.26 \times 10^{-8}$ accumulated in the mass range from $2m_K$ to $m_B-m_D$.
Besides this, a ratio $91.04\%$ for $B^0 \to D^+ \rho(770)^- \to D^+ K^0K^-$
can also be obtained by calculating the corresponding branching ratios in the
ranges [$2m_K$,$2~{\rm GeV}$] and [$2m_K$,$m_B-m_D$].
These indicate that the PQCD predictions for the present processes are reasonable when considering that the current results
still have large uncertainties.

\begin{table}[htb]
\begin{center}
\caption{The comparison of the available experimental measurements for the branching fractions of the $B \to D \rho(770)$ decays and
the PQCD predictions for the branching ratios of the relevant decay modes with the subprocess $\rho(770)\to K \bar{K}$.}
\label{tab3}
\begin{tabular}{ c  c c } \hline\hline
{\rm Decay modes} & {\rm ${\cal B}_{exp}$~\cite{PDG2020}}& {\rm ${\cal B}_{th}$}  \\ \hline
$B^+ \to \bar{D}^0 \rho(770)^+  $&\;\;\;\;\;\;$(1.34\pm0.18)\times10^{-2}$\;\;\;\;\;\;&$ (1.18^{+0.63}_{-0.43})\times10^{-4}$ \\
$B^+ \to D_s^+ \rho(770)^0  $&$<3.0\times10^{-4}$&$ (6.26^{+3.18}_{-1.59})\times10^{-8}$ \\
$B^0 \to D^- \rho(770)^+  $&$(7.6\pm1.2)\times10^{-3}$&$ (7.93^{+5.06}_{-3.01})\times10^{-5}$\\
$ B^0 \to D_s^+ \rho(770)^- $ &$<2.4\times10^{-5}$&$ (2.32^{+1.80}_{-1.15})\times10^{-7}$ \\
$ B^0 \to \bar{D}^0 \rho(770)^0 $&$(3.21\pm0.21)\times10^{-4}$&$ (1.07^{+0.92}_{-0.69})\times10^{-6}$\\
$ B_s^0 \to D_s^- \rho(770)^+   $&$(6.9\pm1.4)\times10^{-3}$&$(6.06^{+3.50}_{-2.11})\times10^{-5} $   \\
 \hline\hline
\end{tabular} \end{center}
\end{table}

For comparison, we list the available experimental measurements for the branching fractions of the two-body
$B \to D \rho(770)$ decays from the {\it Review of Particle Physics}~\cite{PDG2020} in Table~\ref{tab3}, together
with the PQCD predictions for the branching ratios of the relevant decay
modes with the subprocess $\rho(770)\to K \bar{K}$ shown in Tables~\ref{tab1} and~\ref{tab2}.
The ratios between the relevant branching fractions are
 \begin{eqnarray}\label{ratio}
 R_1 &=& \frac{ {\cal B}(B^+ \to \bar{D}^0 \rho(770)^+ \to \bar{D}^0K^+\bar{K}^0)}{{\cal B}(B^+ \to \bar{D}^0 \rho(770)^+)}
= 0.0088^{+0.0048}_{-0.0034}, \nonumber\\
R_2 &=& \frac{ {\cal B}(B^0 \to D^- \rho(770)^+ \to D^- K^+\bar{K}^0)}{{\cal B}(B^0 \to D^- \rho(770)^+)}
=0.010^{+0.007}_{-0.004}, \nonumber\\
R_3 &=& \frac{ {\cal B}(B^0 \to \bar{D}^0 \rho(770)^0 \to \bar{D}^0 K^+K^-)}{{\cal B}(B^0 \to \bar{D}^0 \rho(770)^0)}
=0.0033^{+0.0029}_{-0.0022}, \nonumber\\
R_4 &=& \frac{ {\cal B}(B_s^0 \to D_s^- \rho(770)^+\to D_s^- K^+\bar{K}^0)}{{\cal B}(B_s^0 \to D_s^- \rho(770)^+)}
=0.0088^{+0.0054}_{-0.0035}.
 \end{eqnarray}
Due to the suppression from the phase space, the predicted branching fractions of the quasi-two-body decays
$B^+ \to \bar{D}^0 \rho(770)^+\to \bar{D}^0  K^+\bar{K}^0$, $B^0 \to D^- \rho(770)^+\to D^- K^+\bar{K}^0$, and
$B_s^0 \to D_s^- \rho(770)^+ \to D_s^-  K^+\bar{K}^0$ are around $0.9\%$ of the experimental data for the corresponding
two-body cases, while a ratio near $0.4\%$ for $B^0 \to \bar{D}^0 \rho(770)^0 \to \bar{D}^0 K^+K^-$ is found.

With the relations~\cite{epjc39-41,epjc80-815}
 \begin{eqnarray}
|c_{\rho^0}|\approx\frac{f_{\rho(770)}|g_{\rho(770)^0K^+K^-}|}{\sqrt2 m_{\rho(770)}}, \qquad
|c_{\rho^+}|\approx\frac{f_{\rho(770)}|g_{\rho(770)^+K^+\bar{K}^0}|}{ m_{\rho(770)}},\qquad
|c_{\rho^-}|\approx\frac{f_{\rho(770)}|g_{\rho(770)^-K^0K^-}|}{ m_{\rho(770)}}
\label{eqs-crho}
\end{eqnarray}
and Eq.~(\ref{def-F-K+0}), one can obtain the relation between the strong couplings
$|g_{\rho(770)^+K^+\bar{K}^0}|=|g_{\rho(770)^-K^0K^-}|\approx \sqrt2 |g_{\rho(770)^0K^+K^-}|$, which leads to
$\Gamma_{\rho(770)^+K^+\bar{K}^0}=\Gamma_{\rho(770)^-K^0K^-}\approx 2\Gamma_{\rho(770)^0K^+K^-}$.
When considering $\Gamma_{\rho(770)^\pm \pi^\pm \pi^0}=\Gamma_{\rho(770)^0\pi^+\pi^-}$
and the relation in Eq.~(\ref{quasi-da}), we have
\begin{eqnarray}
  \frac{{\cal B}(B\to D \rho(770)^+ \to DK^+\bar{K}^0)}{{\cal B}(B\to D \rho(770)^+\to \pi^+\pi^0)}
     =\frac{{\cal B}(B\to D \rho(770)^- \to DK^-K^0)}{{\cal B}(B\to D \rho(770)^- \to \pi^-\pi^0)}
     \approx2\frac{{\cal B}(B\to D \rho(770)^0 \to DK^+K^-)}{{\cal B}(B\to D \rho(770)^0 \to \pi^+\pi^-)}.
 \label{R}
\end{eqnarray}
Obviously, the above theoretical analysis is consistent with the numerical results
based on the fact that most of the experimental data were measured by assuming ${\cal B}(\rho(770)\to \pi\pi) \approx 100\%$.
For the branching fractions of decays $B^+ \to D_s^+ \rho(770)^0$ and  $B^0 \to D_s^+ \rho(770)^-$,
 no specific values but the upper limits of $3.0\times10^{-4}$ and $2.4\times10^{-5}$ at a $90\%$ confidence
 level were given by the CLEO and BABAR Collaborations~\cite{prl70-2681,prd78-032205}. Utilizing the PQCD predictions
${\cal B}(B^+\to D_s^+\rho(770)^0)=1.52 \times 10^{-5}$ and ${\cal B}(B^0 \to D_s^+ \rho(770)^-)=2.82 \times 10^{-5}$
taken from our previous work in Ref.~\cite{npb923-54}, and  ${\cal B}(B^+\to D_s^+\rho^0 \to D_s^+K^+K^-)=6.26 \times 10^{-8}$
and ${\cal B}(B^0 \to D_s^+ \rho(770)^- \to D_s^+ K^0K^-)=2.32 \times 10^{-7}$ in this work, ratios around
$0.5\%$ and $1\%$, respectively, can be obtained. Also, from the comparison of the results in Ref.~\cite{npb923-54} and this work,
we can find the similar ratios for other decay channels.
Thus, we estimate the branching fractions
${\cal B}(\rho(770)^+ \to K^+\bar{K}^0)={\cal B}(\rho(770)^- \to K^-K^0)\approx 1\%$ and
${\cal B}(\rho(770)^0 \to K^+K^-)={\cal B}(\rho(770)^0 \to K^0\bar{K}^0)\approx 0.5\%$.
In consideration of the large uncertainties, more precise data from LHCb and Belle-II are expected to test our predictions.

\section{Summary} \label{sec-con}

In this work, we analyzed the contributions for the kaon pair originating from the intermediate state $\rho(770)$ for the three-body
decays $B\to D K\bar{K}$ in the PQCD approach. By the numerical evaluations and the phenomenological analyses, we found the
following points:
\begin{itemize}
\item[(i)]
The decay mode of $B \to D \rho(770)^0$ with the intermediate-state $\rho(770)^0$ decays into $K^0\bar{K}^0$ has the same
branching fraction as the corresponding mode with the subprocess $\rho(770)^0 \to K^+K^-$.

\item[(ii)]
Our predictions for the corresponding branching fractions of the  decay modes with the subprocess $\rho(770)^0 \to K^+K^-$
are much less than the measured branching fractions for the three-body decays $B^0 \to \bar{D}^0 K^+ K^-$,
$B_s^0 \to \bar{D}^0 K^+ K^-$, and $B^+ \to D_s^+ K^+K^-$, while the percentage at about $20\%$ of the total three-body
branching fraction for the quasi-two-body decay $B^+ \to \bar{D}^0 \rho(770)^+ \to \bar{D}^0 K^+ \bar{K}^0$ was predicted in
this work.

\item[(iii)]
The branching ratio for the decay $B^+ \to D_s^+ \rho(770)^0 \to D_s^+ K^+K^-$ predicted in this work is of the same magnitude as
that for  $B^+ \to D_s^+ \phi(1020)^0 \to D_s^+ K^+K^-$ measured by LHCb within large uncertainties.

\item[(iv)]
We estimate the branching fractions ${\cal B}(\rho(770)^+ \to K^+\bar{K}^0)={\cal B}(\rho(770)^- \to K^-K^0)\approx 1\%$ and
${\cal B}(\rho(770)^0 \to K^+K^-)={\cal B}(\rho(770)^0 \to K^0\bar{K}^0)\approx 0.5\%$ by comparing the available experimental
measurements and the PQCD predictions for the branching fractions of the  $B \to D \rho(770)$ decays
with the PQCD predicted branching ratios of the relevant decay modes $B \to D \rho(770) \to D K \bar{K}$ in this work.

\end{itemize}

\begin{acknowledgments}
This work was supported by the National Natural Science Foundation of China under Grants No.~11947011 and
No.~11547038. A.~J.~Ma was also supported by the Natural Science Foundation of Jiangsu Province under Grant
No.~BK20191010 and the Scientific Research Foundation of Nanjing Institute of Technology under Grant No.~YKJ201854.
\end{acknowledgments}

\appendix
\section{DECAY AMPLITUDES} \label{sec-app}
The expressions for amplitudes from diagrams ($a_1$-$d_1$) of Fig.~\ref{fig:fig1}  are written as
\begin{eqnarray}
F_{e\rho}^{LL}&=&8\pi C_F m^4_B f_D\int  dx_B dz
\int  b_B db_B b db  \phi_B  \big\{\big[ [-\bar{\eta} (1+z)+r^2(1+2\bar{\eta})z
-r^4\bar{\eta}z ]\phi_0 -\sqrt{\eta (1-r^2) }\nonumber\\
&&\times[\bar{\eta}(1-2(1-r^2)z)(\phi_s +\phi_t )+r^2(\phi_s -\phi_t )]  \big] E_e(t_a)h_a(x_B,z,b,b_B)S_t(z)
-\big[(1-r^2 )[\eta\bar{\eta}+r^2 (x_B-\eta )  ] \phi_0\nonumber\\
&& +2\sqrt{\eta (1-r^2) }[\bar{\eta}-r^2(1-2\eta+x_B)]\phi_s \big] E_e(t_b)h_b(x_B,z,b_B,b)S_t(|x_B-\eta|)\big\},
\label{eq:f01}
\end{eqnarray}
\begin{eqnarray}
M_{e\rho}^{LL}&=&32\pi C_F m^4_B/\sqrt{6} \int  dx_B dz dx_3
\int  b_B db_B b_3 db_3  \phi_B \phi_D  \big\{\big[[(r^2(r^2-\eta)-\bar{\eta})(\bar{\eta}(1-x_3)-x_B-\eta z)\nonumber\\
&&+r(r_c(\bar{\eta}-r^2)+\eta r(\bar{\eta}+r^2))] \phi_0 +\sqrt{\eta (1-r^2) }[-r^2(x_B+\bar{\eta}x_3)(\phi_s +\phi_t )+
\bar{\eta}(1-r^2)z(\phi_s -\phi_t )+2r(\bar{\eta}r\nonumber\\
&&-2r_c)\phi_s ] \big]E_n(t_c)h_c(x_B,z,x_3,b_B,b_3)+\big[ ( r^2-\bar{\eta})  (x_B-(1-r^2)z-\bar{\eta}x_3 ) \phi_0
+\sqrt{\eta (1-r^2) }[r^2(x_B-\bar{\eta}x_3)\nonumber\\
&&\times (\phi_s -\phi_t ) -\bar{\eta}(1-r^2)z(\phi_s +\phi_t )]\big] E_n(t_d)h_d(x_B,z,x_3,b_B,b_3) \big\}.
\label{eq:m01}
\end{eqnarray}
The expressions for amplitudes from diagrams ($e_1$-$h_1$) of Fig.~\ref{fig:fig1}  are written as
\begin{eqnarray}
 F_{a\rho}^{LL}&=&8\pi C_F m^4_B f_B\int  dx_3 dz
\int  b_3 db_3 b db  \phi_D  \big\{\big[ [-\bar{\eta}(1-r^2)^2z+(1-2r r_c)(\bar{\eta}-r^2)]
\phi_0+\sqrt{\eta (1-r^2) } \nonumber\\
&&\times[r_c\bar{\eta}
(\phi_s +\phi_t )+r(2(1-r^2)z+rr_c)(\phi_s -\phi_t ) -4r\phi_s  ] \big]E_a(t_e)h_e(z,x_3,b,b_3)S_t(z)
+\big[ [(r^2-1 ) ((\bar{\eta}-r^2)\eta \nonumber\\
&& +\bar{\eta}^2x_3) ] \phi_0+2r\sqrt{\eta (1-r^2) }[\bar{\eta}(1+x_3)+2\eta -r^2 ] \phi_s \big]
E_a(t_f)h_f(z,x_3,b_3,b)S_t(|\eta(x_3-1)-x_3|)\big\},
\label{eq:f02}
\end{eqnarray}
\begin{eqnarray}
M_{a\rho}^{LL}&=&32\pi C_F m^4_B/\sqrt{6} \int  dx_B dz dx_3
\int  b_B db_B b db  \phi_B  \phi_D \big\{\big[[\eta\bar{\eta}+ r^2(r^2- 1) +(\bar{\eta}+r^2(\eta- r^2)) \nonumber\\
&&\times(x_B+\eta z+\bar{\eta}x_3)] \phi_0+r\sqrt{\eta (1-r^2) }[((1-z)r^2+z)(\phi_s +\phi_t )+(-x_B+\bar{\eta}(1-x_3))(\phi_s -\phi_t )
-4 \phi_s  ] \big] \nonumber\\
&& \times E_n(t_g)h_g(x_B,z,x_3,b,b_B)+\big[[ (\bar{\eta}- r^2 )(-\bar{\eta}(1-z)+ (-\eta+x_B+\bar{\eta}(1-z-x_3))r^2)] \phi_0
 +r\sqrt{\eta (1-r^2) } \nonumber\\
&&\times[(-x_B-\bar{\eta}(1-x_3))(\phi_s +\phi_t )-(z+(1-z)r^2)(\phi_s -\phi_t )+2\phi_s  ]\big] E_n(t_h)h_h(x_B,z,x_3,b,b_B)
\big\}. \label{eq:m02}
\end{eqnarray}
The expressions for amplitudes from diagrams ($m_1$-$p_1$) of Fig.~\ref{fig:fig1} are written as
\begin{eqnarray}
F_{eD}^{LL}&=&8\pi C_F m^4_B F_K\int  dx_B dx_3
\int  b_B db_B b_3 db_3 \phi_B \phi_D  \big\{ (1+r)  [-\bar{\eta}-x_3+\eta^2(r-1)x_3+2\eta(r-1)^2x_3\nonumber\\
&&+r(-2rx_3+r+3x_3) ]
E_e(t_m)h_m(x_B,x_3,b_3,b_B)S_t(x_3)+\big[\bar{\eta}(r_c+\eta x_B)+2r(-\eta x_B-\bar{\eta}(1+r_c))\nonumber\\
&&+r^2(\bar{\eta}^2-r_c)+2r^3(1+r_c)-\bar{\eta}r^4)\big] E_e(t_n)h_n(x_B,x_3,b_B,b_3)S_t(x_B) \big\}, \label{eq:f03}
\end{eqnarray}
\begin{eqnarray}
 M_{eD}^{LL}&=&32\pi C_F m^4_B/\sqrt{6} \int  dx_B dz dx_3
\int  b_B db_B b db \phi_B \phi_D \phi_0 \big\{\big[-\bar{\eta}^2(1-x_B-z)+rx_3+\eta r(x_B+z-x_3)+\bar{\eta}r^2\nonumber\\
&&\times(\eta(z+x_3-2)-x_B-2z-x_3+2)-r^3(\eta z +\bar{\eta}x_3)-r^4(-\bar{\eta}(z+x_3)-2\eta+1)\big] E_n(t_o)h_o(x_B,z,x_3,b_B,b)\nonumber\\
&&+\big[(r-1)(\bar{\eta}+r)(x_B+(r^2-1)z)+\bar{\eta}(\bar{\eta}-(1+r-r^2)r)x_3 \big]E_n(t_p)h_p(x_B,z,x_3,b_B,b) \big\}. \label{eq:m03}
\end{eqnarray}
The expressions for amplitudes from diagrams ($a_2$-$d_2$) of Fig.~\ref{fig:fig2} are written as
\begin{eqnarray}
F_{e\rho}^{LL}&=&8\pi C_F m^4_B f_D\int  dx_B dz
\int  b_B db_B b db  \phi_B  \big\{\big[ [-\bar{\eta} (1+z)+r^2(1+2\bar{\eta}z)
-r^4\bar{\eta}z ]\phi_0 -\sqrt{\eta (1-r^2) }\nonumber\\
&&\times[\bar{\eta}(1-2(1-r^2)z)(\phi_s +\phi_t )+r^2(\phi_s -\phi_t )]  \big] E_e(t_a)h_a(x_B,z,b,b_B)S_t(z)
-\big[(1-r^2 )[\eta\bar{\eta}+r^2 (x_B-\eta )  ] \phi_0\nonumber\\
&& +2\sqrt{\eta (1-r^2) }[\bar{\eta}-r^2(1-2\eta+x_B)]\phi_s \big] E_e(t_b)h_b(x_B,z,b_B,b)S_t(|x_B-\eta|)\big\},
\end{eqnarray}
\begin{eqnarray}
M_{e\rho}^{LL}&=&32\pi C_F m^4_B/\sqrt{6} \int  dx_B dz dx_3
\int  b_B db_B b_3 db_3 \phi_B \phi_D \big\{\big[[(\bar{\eta} +r^2) (1 -r^2)(x_B+\eta z-\bar{\eta}x_3)] \phi_0\nonumber\\
&& +\sqrt{\eta (1-r^2) }[r^2(-x_B+\bar{\eta}x_3)(\phi_s +\phi_t )+\bar{\eta}(1-r^2)z(\phi_s -\phi_t )  ]  \big]
E_n(t_c)h_c(x_B,z,x_3,b_B,b_3)+\big[(-\bar{\eta}+r^2)\nonumber\\
&&\times[x_B-z+r(r(z-1)+r_c)-\bar{\eta}(1-x_3)]  \phi_0 +\sqrt{\eta (1-r^2) }\big[ -\bar{\eta}(1-r^2)z(\phi_s +\phi_t
)+r^2(\bar{\eta}x_3+x_B)(\phi_s -\phi_t )\nonumber\\
&&+2 (2rr_c-\bar{\eta}r^2)\phi_s ]\big] E_n(t_d)h_d(x_B,z,x_3,b_B,b_3) \big\}.
\end{eqnarray}
The expressions for amplitudes from diagrams ($e_2$-$h_2$) of Figs.~\ref{fig:fig2} are written as
\begin{eqnarray}
 F_{aD}^{LL}&=&8\pi C_F m^4_B f_B\int  dx_3 dz
\int  b_3 db_3 b db \phi_D \big\{\big[  (r^2-1)[\eta(-\bar{\eta}+r^2)-\bar{\eta}^2 x_3]\phi_0
+2 r\sqrt{\eta (1-r^2) }[  1+\eta\nonumber\\
&&+\bar{\eta}x_3-r^2] \phi_s \big] E_a(t_e)h_e(z,x_3,b_3,b)S_t(x_3)-\big[[ \bar{\eta}(r^4(z-1)+r^2(\bar{\eta}-2z)
+z-2rr_c)+2r^3r_c] \phi_0\nonumber\\
&&+\sqrt{\eta (1-r^2) } [r(2z+2r^2(1-z)-rr_c)(\phi_s+\phi_t)+\bar{\eta}(2r-r_c)(\phi_s-\phi_t)]\big] E_a(t_f)h_f(z,x_3,b,b_3)S_t(z) \big\},
\end{eqnarray}
\begin{eqnarray}
 M_{aD}^{LL}&=&32\pi C_F m^4_B/\sqrt{6} \int  dx_B dz dx_3
\int  b_B db_B b db \phi_B \phi_D \big\{\big[(-\bar{\eta}+r^2)[\bar{\eta}(r^2(z-x_3)-x_B-z)+r^2-\eta] \phi_0\nonumber\\
&&+r\sqrt{\eta (1-r^2) }[  (z(1-r^2)+x_B)(\phi_s+\phi_t)+\bar{\eta}x_3(\phi_s-\phi_t)+2\phi_s]  \big]
E_n(t_g)h_g(x_B,z,x_3,b,b_B)-\big[   (\bar{\eta} +r^2 )\nonumber\\
&& \times[(1-r^2)(\bar{\eta}x_3-\eta z)+x_B\eta ] \phi_0+r\sqrt{\eta (1-r^2) }\big[ \bar{\eta}x_3(\phi_s+\phi_t)+((1-r^2)z-x_B)(\phi_s-\phi_t)]\big] \nonumber\\
&&E_n(t_h)h_h(x_B,z,x_3,b,b_B)
\big\}.
\end{eqnarray}
In the formulas above, the symbol $\bar{\eta}=1-\eta$, the mass ratio $r=\frac{m_D}{m_B}$, and $r_c=\frac{m_c}{m_B}$ are
adopted. The values $b_B, b$, and $b_3$ are the conjugate variables of the transverse momenta of the light quarks in the $B$ meson,
resonance $\rho(770)$, and $D$ meson. The explicit expressions for the hard functions $h_i$, the evolution factors $E(t_i)$, and the
threshold resummation factor $S_t$ can be found in Ref.~\cite{npb923-54}.


\end{document}